\documentclass{article}
\usepackage{natbib,emulateapj,latexsym,apjfonts}

\newcommand{\rh}{\rho}
\newcommand{\si}{\sigma}

\newcommand{\Om}{\Omega}

\newcommand{\f}{\frac}

\begin{document}

\title{Strange stars and superbursts at near-Eddington mass accretion rates }

\author{Monika Sinha \altaffilmark{1}}
\affil{CSIR Senior Research Fellow, Department of Physics,
Presidency College, 86/1 College Street, Kolkata 700073, India}

\author{Subharthi Ray \altaffilmark{2}}
\affil{Inter University Centre for Astronomy and Astrophysics
(IUCAA), Post Bag 4, Pune 411007, India}

\author{Mira Dey\altaffilmark{3}}
\affil{ CSIR Emeritus Scientist, Department of Physics, Presidency
College, 86/1 College Street, Kolkata 700073, India }

\and

\author{Jishnu Dey \altaffilmark{3}}
\affil{Department of Physics, Presidency College, 86/1 College
Street, Kolkata 700073, India}

\altaffiltext{1}{email: monika2003@vsnl.net}

\altaffiltext{1}{email: sray@iucaa.ernet.in}

\altaffiltext{3}{Work supported by grant SP/S2/K-03/2001, Dept of
Science and Technology, Govt. of India, email:
deyjm@giascl01.vsnl.net.in}

\begin{abstract}
Careful assessment of four good superburst candidates for
GX~17$+$2 reveals that superburst is possible at near Eddington
mass accretion rates. For the other seven stars, where superburst
is found, there is the standard model of burning accumulated
carbon from normal type I bursts of the accreting stars. However,
there is the need for carbon, nitrogen and oxygen mass fraction
($Z_{CNO}$) which must be larger than $Z_{CNO,\odot}$, where the
latter refers to the standard value found in the sun. Also it is
very difficult to incorporate GX~17$+$2 into the standard picture
of superbursts. In case of superbursts from strange stars,
arising from broken quark pairs going over to diquarks at the
surface of the star, these problems do not arise. Furthermore
there is a natural explanation for the large value of $\sim ~
1000$ for $\alpha$ which is defined in the literature as the
ratio of energy released between normal bursts to the energy
released during the normal burst. In the scenario for superbursts
in strange stars it may be argued that the relatively smaller
value of $\alpha$ of $\sim$ 440 indicates frequent recurrence of
superbursts which is reflected in 4U 1636$-$53.
\end{abstract}


\section{Introduction}

Type I X-ray bursts from Low Mass X-ray Binary (LMXB) systems are
believed to be due to the thermonuclear fusion. The duration of
such bursts is typically of the order of seconds to minutes.
Recently some such Type I X-ray bursters show bursts three orders
of magnitude more energetic also longer by same factor compared to
type I X-ray bursts. That is why they are known as long bursts.
Also for their large fluences they are often referred as
superbursts.

Eight different superbursts have been detected with accretion rate
$\sim (0.1-0.3)\times \dot{M}_{Edd}$ (Eddington mass accretion
rate). They are from 4U~1735$-$44 \cite[]{corn00}, Serpens~X-1
\cite[]{corn02}, GX~3$+$1 \cite[]{kuul01}, KS~1731$-$26
\cite[]{kuul02}, 4U~1820$-$30 \cite[]{sb}, 4U~1636$-$53
\cite[]{rudy01} and 4U~1254$-$69 \cite[]{int}.

All of them show some common typical features, like all
superbursters are known Type-I X-ray bursters; the burst duration
is long $\sim$ few hours; burst energy is large - of the order of
$10^{42}$ ergs; there exists a persistent preburst luminosity
$(0.1-0.3) \times L_{Edd}$. It was believed that there are no
super burst from sources with persistent luminosity less than $0.1
L_{Edd}$ and greater than $0.5 L_{Edd}$. However four superbursts
were reported from GX 17$+$2 in March 2004 \cite[]{int}. One of
the seven superbursters, $4U~1636-53$, has shown repeated
superbursts in a time interval 4.7 years \cite[]{rudy01}.
in't~Zand et al. (2004) chose a star with near Eddington mass
accretion rate, GX 17$+$2,  and found at least four superbursts
with mean recurrence time $30\pm15~d$. The shortest observed
recurrence time is $8.2~d$.

The characteristics of the seven superbursters known so far are
given in Table (\ref{chc}). We have excluded the source GX~17$+$2
here.

\begin{table*}[htbp]
\caption{Properties of superbursters. $L_{Peak}$ is peak
luminosity, $E_b$ is total fluence, $L_{per}$ is persistent
luminosity and $\tau_{exp}$ is decay time.} \vskip 1cm
\begin{center}
\begin{tabular}{|c|c|c|c|c|c|c|c|}
\hline
Stars&$4U~1735$&$Serpens$&$KS~1731$&$4U~1636$&$4U~1820$&$GX$&$4U~1254$
\\ &$-44$&$X-1$&$-260$&$-53$&$-30$&$3+1$&$-690$\\ \hline
Distances&$9.2$&$8.4$&$7$&$5.9$&$7.6$&$4-6$&$13\pm 3$\\
(kpc)&&&&&&&\\ \hline
Duration&$\sim7$&$\sim4$&$\sim12$&$\geq1-3$&$\sim
3$&$4.4-16.2$&$14\pm2$\\ (Hrs.)&&&&&&&\\ \hline Energy &&&&&&&\\
range&$2-28$&$2-28$&$2-28$&$1.5-12$&$2-60$&$1.5-12$&$2-28$\\
(keV)&&&&&&&\\ \hline
$L_{Peak}$&$1.05\pm0.1$&$1.6\pm0.2$&$1.4\pm0.1$&$\sim1.2$&$3.4\pm0.1$&$\sim0.5$&$.44\pm0.2$\\
($10^{38}erg~s^{-1}$)&&&&&&&\\ \hline
$E_b$&$\sim0.5$&$\sim0.8$&$\sim1.0$&$\sim0.5-1.0$&$\geq1.4$&$\sim0.5-2.0$&$0.8\pm0.2$\\
($10^{42}ergs$)&&&&&&&\\ \hline
$L_{per}$&$\sim0.25$&$\sim0.2$&$\sim0.1$&$\sim0.1$&$\sim0.1$&$\sim0.2$&$\sim0.13$\\
($L_{Edd}$)&&&&&&&\\ \hline
$kT_{max}$&$\sim2.6\pm0.2$&$\sim2.6\pm0.2$&$\sim2.4\pm0.1$&$-$&$\sim3.0$&
$\sim1.0-2.0$&$1.8\pm0.1$\\ (keV)&&&&&&&\\ \hline
$\tau_{exp}$&$1.4\pm0.1$&$1.2\pm0.1$&$2.7\pm0.1$&$1.5\pm0.1$&$\sim1.0$&$\sim1.6$&$6\pm0.3$\\
(Hrs.)&&&&&&&\\
 \hline
  $\alpha$&4400&5800&780&440&2200&2100&4800\\
 \hline
\end{tabular}
\end{center}
\label{chc}
\end{table*}
We have used the strange star model proposed by  Dey et al. (1998)
and which has been used by Li et al. (1999a) for compactness and
Li et al. (1999b) and Mukhopadhyay et. al. (2003) for
quasi-periodic oscillations in the power spectrum. This model
uses an interquark potential which has asymptotic freedom,
confinement - deconfinement transition built into it and uses
density dependent quark mass. The beta stability and charge
neutrality demand a self consistent calculation of the chemical
potentials of the quarks and electrons since interquark
interaction is present and also involves contribution from
density dependence of quark mass an gluon screening length.
Diquark pairing and pair breaking in this model has been used as
a source  to explain superbursts \cite []{mnras}.

The standard explanation of superbursts for neutron star
candidates are in the form of carbon burning. This has been
studied in detail (e.g., by Cooper et. al., 2005) where some
earlier references may also be found. This recent work states that
clearly the observations of $GX ~17+2$ are inconsistent with
their results but that the issue should be investigated further.
These authors also leave out the system $4U~1820-303$ in which
the accreted matter is dominated by helium but point out that it
is not understood theoretically why this system does not exhibit
normal bursts for long periods when the accretion is near its
maximum. As already mentioned a further problem is that the
($Z_{CNO}$) must be larger than $Z_{CNO, \odot}$. It may be
however interesting to see what one gets for the carbon burning
model using the Dey et al. (1998) strange star equation of state
(EOS) for the quark matter. There, the EOS parameters are
determined from hadronic energy and magnetic moment calculations
\cite[]{bagchi}.


In our earlier paper \cite[]{mnras} we put forward the idea of
quark pairing as a possible source of superbursts qualitatively.
In this work  we calculate the details quantitatively and show
that our strange star EOS along with quark pairing could indeed be
the source of superbursts. This is as given in the following
section along with numerical results. In particular we indicate an
alternate scenario where in the strange star with carbon burning
from scenario, the heating mechanism could be due to diquark
pairing rather than transformation of normal accreted matter to
strange matter.

\section{Pair Formation}

We get a spin-spin potential between quarks in specific colour
channels with a smeared Gaussian potential with a renormalized
strength. The smearing and the strength can be obtained by fitting
them to observables like nucleon$-\Delta $ mass splitting  and the
magnetic dipole transition from $\Delta $ to nucleon. We borrow
the allowed sets from Dey \& Dey (1984). The form of the potential
is
\begin{equation}
{\rm H}_{ij}~=~- \f{2\alpha_s\sigma^3}{3 m_i m_j
\pi^{1/2}}(\lambda_i.\lambda_j)({\rm S}_i.{\rm S}_j)e^{-\sigma^2
r_{ij}^2}. \label{diq}
\end{equation}
The factor $\sigma^3/\pi^{1/2}$ normalizes the potential. In this
equation $\alpha_s$ is the strong coupling constant, and the
subscripted $m, \lambda$ and $S$ are the constituent masses,
colour matrices and spin matrices for the respective quarks. For
{\it u-d} quarks this gives $\sigma$ varying from 6 to 2.03
$fm^{-1}$ for a set of $\alpha_s$ 0.5 to 1.12 \cite[]{dd}. It is
found that diquark binding depends strongly on the strength and
range of spin-spin interaction which are interconnected via
hadron phenomenology. This is irrespective of whether it is
deduced from a the Fermi-Breit interquark force or an instanton -
like four fermion interaction \cite[]{raja}.

For colour symmetric state ($6$) $\lambda_i.\lambda_j~=~\f 43$,
colour antisymmetric  ($\bar 3$) $\lambda_i.\lambda_j~=~-\f 83 $
and for  spin symmetric state (triplet) $S_i.S_j~=~\f 14$ (b) spin
antisymmetric (singlet) $S_i.S_j~=~-\f 34 .$

The spin and colour factor gives for flavor symmetric state,
either $\f 14~\times~ \left(-\f 83\right)~=~-\f 23$, or $\left(-\f
34\right)~\times~ \f 43~=~-1.$ So in both cases the potential is
repulsive. And for flavor antisymmetric state, either $\f
14~\times~ \f 43~=~\f 13,$ or, $\left(-\f 34\right)~\times~
\left(-\f 83\right)~=~2,$ showing the potential to be attractive.

The expectation value of the potential (Eq. (\ref{diq})) is taken
between two two-body free particle wave functions $|ij>$ given by
\begin{equation} |ij>= \f{e^{-i{\bf k}_i.{\bf r}_i}}{\sqrt\Om}\f{e^{-i{\bf
k}_j.{\bf r}_j}}{\sqrt\Om}= \f1\Om e^{-i{\bf k}_i.{\bf r}_i}
e^{-i{\bf k}_j.{\bf r}_j}
\end{equation}
where $\Om~=~ \f N\rh $ is the total volume, N being the total
number of quarks and $\rh$ the total quark number density. Now
\begin{equation}
e^{-i{\bf k}_i.{\bf r}_i} e^{-i{\bf k}_j.{\bf r}_j}=e^{-i{\bf
k}.{\bf r}} e^{-i{\bf P}.{\bf R}}
\end{equation}
with $\bf R$ and $\bf r$ the  centre of mass and relative
coordinates with appropriate momentum conjugates ${\bf P}= {\bf
k}_i+{\bf k}_j$ and ${\bf k}= ({\bf k}_i-{\bf k}_j)/2$. The energy
expectation value is
\begin{eqnarray}
\nonumber
<ji|{\rm H}_{ij}|ij>=- (\lambda_i.\lambda_j)({\rm S}_i.{\rm S}_j)
\f{2\alpha_s\sigma^3}{3 m_i m_j \pi^{1/2}}\\
\f{1}{\Om^2}\left<e^{i{\bf k}.{\bf r}} e^{i{\bf P}.{\bf
R}}\left|e^{-\sigma^2 r^2}\right|e^{-i{\bf k}.{\bf r}} e^{-i{\bf
P}.{\bf R}}\right>.
\label{expectation}
\end{eqnarray}

Integrating over volume this reduces to

\begin{equation} <ji|{\rm H}_{ij}|ij>~=~-(\lambda_i.\lambda_j)({\rm
S}_i.{\rm S}_j) \f{2\alpha_s\sigma^2\pi}{3 m_i m_j \Om}f(k)
\label{mom} \end{equation} where $f(k)=(1-e^{k^2/\si^2})/{k^2}$
and $ k^2=\f{1}{4}[k_i^2 + k_j^2-2 k_i k_j cos(\theta)]$, $\theta$
being the angle between ${\bf k}_i$ and ${\bf k}_j$.

For each particle in pair the energy lowering due to one pair can
be taken as
\begin{eqnarray}
\nonumber
E_p(k)&=&\f12<ji|{\rm H}_{ij}|ij>\\
&=&-\f12(\lambda_i.\lambda_j)({\rm S}_i.{\rm S}_j)
\f{2\alpha_s\sigma^2\pi}{3 m_i m_j \Om}f(k).
\end{eqnarray}
If we consider one $u$ quark, then the lowering in energy of the
$u$ quark will be due to all pair it can form with all $d$ quarks
in the particular spin colour channel for which lowering is
maximum that is spin-singlet and colour antisymmetric channel. So
total lowering of energy of one $u$ quark with momentum $k_u$ is
\begin{equation}
E_u(k_u)~=~\f{\Om}{(2\pi)^3}\int E_p({\bf k}_u,{\bf k}_d)d^3{\bf
k}_d.
\end{equation}

Then average energy lowering of $u$ quark is
\begin{eqnarray}
\nonumber
\overline{E_u}&=&\f{\f{6\Om}{(2\pi)^3}\int E_u(k_u)d^3{\bf
k}_u}{N_u}\\
&=&-\f12(\lambda_i.\lambda_j)({\rm S}_i.{\rm S}_j)
\f{2\alpha_s\sigma^2\pi}{3 m_i m_j n_u}\f{6\times2}{(2\pi)^4}I.
\end{eqnarray}
\noindent where
\begin{equation}
I=\int_0^{k_{fu}}\int_0^{k_{fd}}\int_{-1}^1 f(k_u,k_d,\theta)
k_d^2 dk_d k_u^2 dk_u d(cos\theta_d) \label{dipot}
\end{equation}

\noindent Similarly,
\begin{equation} \overline{E_d}=\f12(\lambda_i.\lambda_j)({\rm S}_i.{\rm S}_j)
\f{2\alpha_s\sigma^2\pi}{3 m_i m_j n_d}\f{6\times2}{(2\pi)^4}I.
\end{equation}
\noindent Hence lowering of one $ud$ pair is
\begin{eqnarray}
\nonumber \overline{E}&=&\overline{E_u}+\overline{E_d}\\
\nonumber &=&-\f12(\lambda_i.\lambda_j)({\rm S}_i.{\rm S}_j)
\f{2\alpha_s\sigma^2\pi}{3 m_i
m_j}\f{n_u+n_d}{n_un_d}\f{6\times2}{(2\pi)^4}I.
\end{eqnarray}

Integrating Eq. (\ref{dipot}) in the range from 0 to $k_f$-s of
respective quarks the contribution of a $ud$ diquark in the
energy has been shown in the Table \ref{corr} for three different
EOSs given by Dey et al. (1998) and for the two possible
colour-spin channels with appropriate parameters $\sigma$ and
$\alpha_s$.
\begin{table*}[t]
\caption{Integrated values for the pairing energy Eq.(\ref{diq})
for different pairs for spin singlet (colour $\bar 3$) states in
$MeV$. For spin triplet (colour $6$) state the energies is six
times less. }
\begin{center}
\begin{tabular}{|c|c|c|c|c|c|c|}
\hline
Colour-&Sets & $\alpha_s$ & $\sigma$& \multicolumn{3}{c|}{EOS}\\
\cline{5-7}
spin&&&$fm^{-1}$ & eos1&eos2&eos3\\
states&&&&&&\\ \hline
spin    &1& 0.5  & 6.0 &-23.578&-23.744&-25.901 \\
singlet &2& 0.5  & 4.56&-23.287&-23.451&-25.561 \\
and     &3& 0.87 & 6.0 &-41.025&-41.316&-45.067 \\
colour  &4& 0.87 & 2.61&-38.225&-38.484&-41.807 \\
$\bar 3$&5& 1.12 & 6.0 &-52.814&-53.188&-58.018 \\
        &6& 1.12 & 2.03&-46.636&-46.941&-50.847 \\
\hline spin   &1& 0.5  & 6.0 &-3.930&-3.957&-4.317 \\ triplet&2&
0.5  & 4.56&-3.881&-3.908&-4.260 \\ and    &3& 0.87 & 6.0
&-6.837&-6.886&-7.511 \\ colour &4& 0.87 & 2.61&-6.371&-6.414&-6.
968 \\ $6$    &5& 1.12 & 6.0 &-8.802&-8.865&-9.670 \\
       &6& 1.12 & 2.03&-7.773&-7.823&-8.474 \\
\hline
\end{tabular}
\end{center}
\label{corr}
\end{table*}

The Table (\ref{corr}) shows that the variation of the correlation
energy is significant, when different sets for the smearing in the
spin-spin potential are chosen. The variation with the EOSs is
comparatively unimportant. We re-emphasize that the maximum
numbers in this table $\sim 50~MeV$ are close to the energy
obtained from conversion of normal accreted matter to strange
matter, if any, and could thus provide the energy needed for
reaching a high temperature as in carbon burning scenario.

\section{Observation and Conclusion}

It is seen from the Table (\ref{corr}) the lowering in energy per
pair due to spin-spin interaction varies and is larger for the
singlet channel where it may be 58 $MeV$ whereas in the triplet
channel it is less than 10. When a Type I burst occurs both
channels may be excited equally and some 30 $MeV$ may be absorbed
for an average pair breaking.

The number density of quarks on the surface of the strange star is
$\sim 0.27~fm^{-3}$. With this number density the re-alignment of
quarks within a depth of a few micron will liberate a total
energy of the order of $10^{47} MeV$.

After a type I burst the broken pairs are converted to diquarks
and there will be an energy release which will couple to the
gravitational energy of the accreted particles and this will lead
to a large alpha between bursts.

The scenario for superbursts from strange stars is that a
micron-thick layer of diquarks, at the star surface, get broken
due to repeated Type I bursts during high accretion. Some of the
broken pairs form diquarks again - thus supplementing the energy
released during accretion between bursts - and producing a large
value of $\alpha$. When the number of broken pairs and diquarks
reach a critical balance, the transition is expected to go like an
avalanche. This is due to the two-fermion to boson transformation
where the number of bosons enhance a transition. For this
phenomenon to take place the accretion must be high. And for
GX~17$+$2 - where the accretion is near the Eddington limit - one
would naturally expect recurring superbursts with short intervals
between them as is indeed observed. After the superbursts, the
reverse process takes over during the subsequent normal bursts,
if any. The crucial fact is that the recombination time scale is
long, since the strong interaction pairing is supplemented by beta
equilibrium and charge neutralization which are slower weak
electromagnetic process. The least observed time interval between
two superbursts is 8.2 days which is 100 times the burst period of
about 2 hours. The ratio is like the ratio of strong to
electromagnetic interaction, showing that pair breaking takes
that much more time, as expected. The values of $\alpha$ find a
natural explanation in the enhanced gravitational field of the
strange star and in the energy released due to pairing of diquarks
broken up during a Type I burst.

We expect that when alpha is large it will signify most of the
broken pairs have changed into diquarks. When alpha is not so
large superburst recurrence is perhaps more likely and we find
this in the case of $4U~1636-53$ where alpha is only 440 and
three superbursts have already been spotted.

\acknowledgements {JD and MD acknowledge illuminating discussion
with B. Mukhopadhyay.}


\begin{thebibliography}{}

\bibitem[Bagchi et. al. 2004]{bagchi}  Bagchi M., Dey M., Daw S. \& Dey J.
2004, \nphysa, 740, 109.

\bibitem[Cornelisse et. al. 2000]{corn00} Cornelisse R., Heise J., Kuulkers E., Verbunt
F. \& in 't Zand J. J. M. 2000, \aap, 357, L21.

\bibitem[Cornelisse et. al. 2002]{corn02} Cornelisse R., Kuulkers E., in 't Zand
J. J. M., Verbunt F. \& Heise J. 2002, \aap, 382, 174.

\bibitem[Cooper et al. 2005]{cmsn} Cooper R. L., Mukhopadhyay B., Steeghs D. \& Narayan
R., ``On the production and survival of carbon fuel for
superbursts on accreting neutron stars: Implications for mass
donor evolution"; astro-ph/0508194 v1.

\bibitem[Dey and Dey 1984]{dd} Dey J. \& Dey M. 1984, Phys. Lett.,  B138, 200.


\bibitem[Dey et. al. 1998]{d98} Dey M., Bombaci I., Dey J., Ray S. \&
Samanta B. C. 1998, Phys.  Lett., B438, 123 ; Addendum 1999, B447,
352 ; Erratum 1999 B467, 303.

\bibitem[in 't Zand et. al. 2003]{int} in 't Zand J. J. M., Kuulkers E., Verbunt F.,
Heise J. \& Cornelisse R. 2003, \aap, 411, L487.

\bibitem[in 't Zand et. al. 2004]{intz} in 't Zand J. J. M., Cornelisse R. \& Cumming A. 2004, \aap, 426, 257.

\bibitem[Kuulkers et. al. 2001]{kuul01} Kuulkers E. 2002, \aap, 383, L5.

\bibitem[Kuulkers et. al. 2002]{kuul02} Kuulkers  E., in 't Zand J. M. M., van
Kerkwijk M. H., Cornelisse R., Smith D. A., Heise J., Bazzano A.,
 CocchiM., Natalucci L. \& Ubertini P. 2002, \aap,
382, 503.

\bibitem[Li et. al. 1999a]{Li1999a}  Li, X. D., Bombaci I., Dey M., Dey J. \&  van
den Heuvel E. P. J. 1999a, \prl, 83  3776.

\bibitem [Li et. al. 1999b]{Li1999b} Li X., Ray S., Dey J., Dey M. \& Bombaci I.
1999b, \apj, 527, L51.

\bibitem[Mukhopadhyay et. al. 2003]{mrdd} Mukhopadhyay B., Ray S., Dey J. \& Dey M. 2003,
\apj, 584, L83.


\bibitem[Rajagopal and Wilczek 2000]{raja} Rajagopal K. \& Wilczek F. 2000, `The Condensed
Matter Physics of QCD', Chapter 35 in the Festschrift in honor of
B. L. Ioffe, ``At the Frontier of Particle Physics / Handbook of
QCD", M. Shifman, ed., (World Scientific). ; hep-ph/0011333.

\bibitem[Strohmayer and Brown 2002]{sb} Strohmayer T. E. \& Brown E. F. 2002,
\apj, 566, 1045.

\bibitem[Sinha et. al. 2002]{mnras} Sinha M., Dey M., Ray S. \& Dey
J. 2002, \mnras, 337, 1368.

\bibitem[Wijnands 2001]{rudy01} Wijnands R. 2001, \apj, 554, L59.

\end{thebibliography}
\end{document}